\documentstyle[11pt,paspconf,psfig]{article}

\def\be{\begin{equation}} \def\ee{\end{equation}}
\def\se#1{\S\ref{sec:#1}}
  
\def\fig#1{Fig.~\ref{fig:#1}}

 \def\cl{\centerline}
 
\def\etal{{\it et al.\ }}  \def\etalns{{\it et al.}}

\def\ltsima{$\; \buildrel < \over \sim \;$} \def\lsim{\lower.5ex\hbox{\ltsima}}
\def\gtsima{$\; \buildrel > \over \sim \;$} \def\gsim{\lower.5ex\hbox{\gtsima}}

\def\ifm#1{\relax\ifmmode#1\else$\mathsurround=0pt #1$\fi}
\def\kms{\,{\rm km\,s\ifm{^{-1}}}}
\def\hmpc{\,h\ifm{^{-1}}{\rm Mpc}}

\def\la{\langle} \def\ra{\rangle}

\def\sss{\scriptscriptstyle}
\def\del{\delta}
\def\delg{\delta_{\rm g}}
\def\sb{\sigma_{\rm b}}
  
\def\betai{\beta_{\sss \rm IRAS}}

\def\solar{\ifmmode_{\mathord\odot}\;\else$_{\mathord\odot}\;$\fi}
\def\msun{{\rm M}_{\solar}}

 \def\omm{\Omega_{\rm m}} 
\def\oml{\Omega_\Lambda} 

\def\pmb#1{\setbox0=\hbox{#1}
 \kern.05em\copy0\kern-\wd0 \kern-.025em\raise.0433em\box0}

 \def\vv{\pmb{$v$}} \def\vx{\pmb{$x$}}
\def\vnabla{\pmb{$\nabla$}}
\def\div{\vnabla\!\cdot\!} 

\def\divv{\div\vv}

\def\bh{\hat b}
\def\bt{\tilde b}
\def\bv{b_{\rm var}}
\def\av#1{\la #1 \ra}

\def\tv{tv}

\begin{document}

\title{Cosmic Flows 99: Conference Summary}
\author{Avishai Dekel}
\affil{Racah Institute of Physics, The Hebrew University, Jerusalem
91904, Israel}

\begin{abstract}
I address the following issues:
All bulk velocity measurements (but one)
are consistent with our standard gravitational instability theory.
New accurate data and reconstruction methods
allow high-resolution dynamical analysis nearby,
revealing Virgo, Ursa Major and Fornax as attractors.
Large peculiar-velocity surveys enable robust reconstruction 
of the dynamical fields on the Great-Attractor scale. 
A decomposition of the velocity field into its local and tidal
components indicates the presence of big perturbations further away.
Cluster velocities start exploring very large scales, revealing
Coma, Shapely and other mass enhancements, and constraining a possible
local Hubble bubble.
Supernovae type Ia (SNIa) are very promising for cosmic flow analysis.
Peculiar velocities do provide unique valuable constraints on cosmological
parameters, e.g., $0.3<\!\omm\!<1$ (95\% confidence) independent of biasing.
Jointly with other data they can confine other parameters such as
$\oml$, $h$, $\sigma_8$, $n$, and the biasing.
Nontrivial features of the biasing scheme can explain much of the span 
of estimates for $\beta$.
Quantitative error analysis is essential in our maturing field;
every method ought to be calibrated with suitable mock catalogs, that are
offered as benchmarks.
\end{abstract}


\section{Introduction}
\label{sec:intro}

This is not a comprehensive review of the conference, but rather a collection
of concluding remarks on some of the central issues which I wish to highlight.
The distinctive feature of this conference was the exposure of several new
observational surveys of peculiar velocities, listed in Table 1.
These data enable dynamical studies in three zones: 
our $30\hmpc$ local neighborhood at high-resolution,
within $\sim\!60\hmpc$ with $\sim\!10\hmpc$ smoothing,
and out to $\sim\!120\hmpc$ at low-resolution.
I use some of the analysis tools developed by my colleagues and myself, 
including POTENT, Wiener Filter (WF), decomposition and 
likelihood analysis, to illustrate some of the potential of these data.
I then address the implications to cosmology and galaxy formation,
and my views of how further progress is ought to be made.

The outline is as follows:
\se{bulk} addresses bulk velocities.
\se{sbf} discusses high-resolution analysis in the local neighborhood.
\se{gapp} reviews the robust analysis in the Great Attractor
          vicinity. 
\se{deco} proposes a decomposition
          of the velocity field into divergent and tidal components.
\se{vls} demonstrates the potential of dynamical
         analysis on very large scales.
\se{cospar} addressed the constraints on cosmological parameters.
\se{bias} evaluates the effect of nontrivial biasing on the range
          of estimates for $\beta$.
\se{mock} stresses the importance of error analysis via mock catalogs.
\se{conc} summarizes my main points.

\def\LA{$L_{\rm m}$-$\alpha$}
\bigskip
\cl{
\begin{tabular}{lcccccl}
\multicolumn{7}{l}{ {\bf Table 1:} Peculiar Velocity Data } \\
\hline\hline
Catalog &Dist. &Err  &Objects &Num.    &Rad.     &Reference \\
        &Ind.  &\%   &        &ga/cl  &$h^{\!-1}\!$Mpc& \\
\hline\hline
SBF       &SBF   & 8 &E/ga    & 300    &30(40)   &Blakeslee \etalns, \tv\\
PT        &TF    &18 &S/ga    & 500    &30(40)   &Pierce, Tully, \tv\\
\hline
ENEAR     &FP    &20 &E/ga,cl &1900    &40(70)   &Wegner \etalns, \tv\\
M3        &TF,FP &18 &S,E/ga,cl &3400  &50(80)   &Willick \etal 97a\\
SFI       &TF    &18 &S/ga    &1650    &50(70)   &Haynes \etal 98a,b\\
Shellflow &TF    &18 &S/ga    & 300    &40-75     &Courteau \etalns, \tv\\
\hline
SNIa      &SN    & 8 &S        &  44    &50(200)  &Riess, \tv \\
\hline
SCI+II    &TF    &18 &S/cl     &1300/76 &95(200) &Dale \& Giovanelli, \tv\\
SMAC      &FP    &20 &E/cl     & 700/56 &65(140)  &Smith \etalns, \tv\\
LP10K     &TF    &18 &S/cl     & 170/15 &90-135    &Willick, \tv\\
BCG       &\LA   &18 &E/cl     & 120    &85(150)  &Lauer \& Postman 94\\
EFAR      &FP    &20 &E/cl     & 450/85 &60-150    &Colless \etalns, \tv\\
\hline\hline
\end{tabular}
}

\section{Bulk Velocity}
\label{sec:bulk}

The simplest quantity extracted from a peculiar
velocity sample is the bulk velocity $V$, in a sphere (or a shell)
about the Local Group (LG). 
The measurements are sometimes referred to as either proving 
``convergence" to the cosmic frame within a given radius, or as 
posing a challenge to the large-scale isotropy of the universe.
I would like to stress that the interpretation
of a bulk velocity is meaningful only in the context of 
a specific theoretical model, and is a quantitative issue.
In fact, large-scale isotropy does not require ``convergence" on 
any finite scale. 
Our models predict, quite robustly, a relatively weak descent of
amplitude with scale, and the large cosmic variance due to the finite,
sparse and nonuniform sampling can accommodate a large range of results.

\begin{figure}
\vskip 7.3cm
{\includegraphics{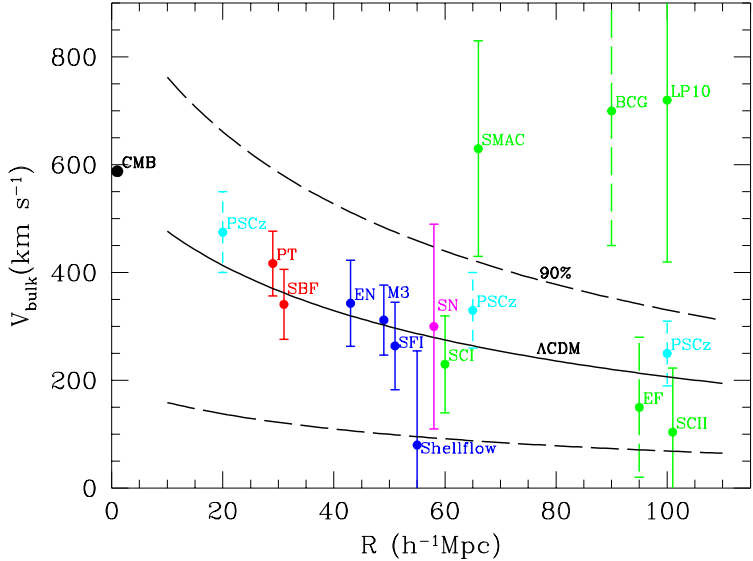}}
\caption{
Amplitude of CMB bulk velocity in top-hat spheres about the LG,
in comparison with theory.
The curves are the predicted rms and cosmic scatter
for a $\Lambda$CDM model.
The measurements, based on the data listed in Table 1,
are crudely translated to a top-hat bulk velocity.
The errorbars are random only.
All the non-zero vectors (except BCG) point to
$(l,b)=(280,0)\pm30^\circ$.
Shown as well are the LG dipole velocity
and linear estimates from the PSCz redshift survey for $\beta=0.7$.
}
\label{fig:bu}
\end{figure}

I therefore point first, in \fig{bu}, to the theoretical prediction 
of a $\Lambda$CDM model for the simplest statistic: the bulk-flow 
amplitude in a top-hat sphere.  The solid line is the rms value,
obtained by integrating over the power spectrum times
the square of the Fourier Transform of the top-hat window.
The dashed lines represent 90\% cosmic scatter in the Maxwellian
distribution of $V$, when only one random sphere is sampled.
This model (flat, with $\omm=0.35$, $n=1$, $h=0.65$), has
$\sigma_8 \omm^{0.5} = 0.51$, consistent with the constraints from
cluster abundance (Eke, Cole \& Frenk 1996). 
In fact, any model from the CDM family that
is normalized in a similar way predicts bulk velocities in the same ball-park,
so the theoretical curves should be regarded as representative
of our ``standard" models.
Note the gradual descent and the large scatter: the velocity could 
almost vanish inside $50\hmpc$, or be as high as $400\kms$ near $100\hmpc$, 
without violating standard cosmology.

A bulk velocity can be computed by fitting a 3D model of 
constant velocity to the observed radial peculiar velocities. 
Each datum contributes to the fit, usually weighted by the inverse
square of the relative distance error, added in quadrature to a constant
velocity dispersion.  Thus, the result corresponds to
a nonuniform window in space, which is typically biased towards 
small radii and is very specific to the sample.
A proper comparison with theory should take into account the sampling
window and the associated cosmic scatter
(Kaiser 1988; Watkins \& Feldman 1995).
However, a semi-quantitative impression can be obtained by a crude comparison
in the ``theory plane", for which one can approximate a top-hat window 
by equal-volume weighting at the expense of larger random errors.
A full POTENT analysis is a more accurate way of mimicking uniform weighting.

The results are put together in a crude way in \fig{bu}, 
displaying the amplitudes of the bulk velocities in the CMB frame, 
as if they all represent top-hat bulk velocities.
The amplitudes can be compared because the directions of all the 
nonzero vectors are remarkably similar: with the exception of BCG, 
they all lie in the $30^\circ$ vicinity of $(l,b)=(280,0)$.
Some of the error-bars are based on a careful error analysis using
mock catalogs, while others are crude estimates. In most cases they 
represent random errors only and underestimate the systematic biases.
The bulk velocities were de-biased by subtracting in quadrature the
errors in each component.

Also shown in \fig{bu} is the velocity of the LG as deduced from the
CMB dipole, and the velocities predicted from the IRAS PSCz redshift
survey using linear theory, with $\betai=0.7$, the best-fit
to the CMB dipole (Saunders \etalns, this volume, hereafter \tv).

M3, SFI and Shellflow are dominated by Tully-Fisher (TF) 
spirals inside $R\lsim 50\hmpc$. 
The M3 result refers to the VM2 calibration (Dekel \etal 1999) and is
a bit lower than the original M3 result.
The M3 and SFI results were obtained via a uniform POTENT reconstruction
and error analysis.  Inside $R\lsim 50\hmpc$ they generally agree, while
at larger radii the bulk velocity in SFI drops faster than in M3.  
This difference may be related to a difference in matching the zero
points of the TF relations between North and South in the two catalogs,
but one should admit that these two samples are not large enough for a
reliable estimate of $V$ beyond $50\hmpc$, where, for one thing,
the Malmquist-bias corrections are quite uncertain.
The new Shellflow result seems to favor a low value, but 
it is for a shell outside the main body of M3 and SFI, and the large error
due to the relatively small number of galaxies makes it consistent with
the model, and with both M3 and SFI, at the $\sim\!1\sigma$ level.
However, the Shellflow data will enable a revised calibration of M3 and
SFI, which can significantly reduce the uncertainties.
The preliminary report from the ENEAR survey of Fundamental-Plane (FP)
velocities (Wegner \etalns, \tv) agrees well with M3 and SFI.

In our local $30\hmpc$ vicinity, we have computed the bulk flows via a minimum
$\chi^2$ fit with volume weights for the two independent new surveys:
the accurate SBF measurements of 300 ellipticals (Tonry \etalns, \tv), 
and TF measurements of 500 spirals (Pierce, Tully, \tv).
A dispersion of $\sim\!300\kms$ is assumed in the fit, 
to make $\chi^2\!\simeq\!{\rm d.o.f}$.
One can see that all the results within $50\hmpc$ are remarkably 
consistent with our theoretical expectations and with each other.

On larger scales we have several new results based on clusters of galaxies:
SMAC and EFAR of FP ellipticals, and LP10K and SCI+II based on TF spirals. 
The EFAR sample is an exception because it covers limited
areas of sky, largely perpendicular to the direction of the bulk flow.
The fact that all these results are consistent with the same bulk-flow 
direction is very comforting in view of the worries raised earlier by 
the BCG result.
 
The amplitudes, on the other hand, show large scatter.
The results are as reported by the observing teams, with
an effective top-hat radius crudely assigned.
The main point is that no one single measurement 
is more than $\sim\!2\sigma$ away from the model prediction,
even in the simplified presentation of in \fig{bu}. 
This is confirmed by a more accurate analysis which
takes into account the systematic errors due to sampling together 
with the random errors and cosmic variance (Hudson, \tv; Hoffman, \tv).
A model with a steeper drop in the power spectrum on the ``blue"
side of the peak, like CHDM, gives a somewhat higher amplitude
and therefore a better fit to SMAC and LP10K.

Hudson demonstrates further that the bulk velocity vectors as measured
in all of these large-scale surveys (except BCG) are in fact consistent with 
each other at the 95\% CL. 
Take for example the ``high" LP10K value compared to the ``low"
SCII value. We note that the individual peculiar velocities of the 7 clusters
common to these samples are consistent within the errors, and that
the 15 SCII clusters that lie within the LP10K shell have a nominal bulk
velocity of $\sim\!400$, closer to the LP10K result.
I therefore do not see the need or justification for Willick (\tv)
to discard his own result; it is high, but consistent with the model
and the other data given the expected (big) errors.

As pointed out by Strauss (\tv), there is no clear understanding yet of
the source of the discrepant BCG result,
and we are therefore eagerly waiting for the upcoming, follow-up, larger BCG
survey for a possible resolution of this mystery.
 
The bulk velocity of SNIa is computed by us, volume weighted and de-biased,
from the 16 SNe inside $60\hmpc$ (out of the sample of 44 inside $300\hmpc$,
Riess, \tv). Even slightly higher values are obtained (with no volume
weights) inside $\sim\!100\hmpc$.
The SN result still carries a large error
because of the small number of objects in the current sample, but
the accurate distances and the unlimited sampling potential
promise that this distance indicator will eventually become very valuable in 
reducing the uncertainties on large scales.

Despite the apparent scatter on large scales,
and the disputes over the bulk velocity being small or large,
we see no significant discrepancies between the bulk velocity
data and models, and thus the bulk flow measurements do not 
introduce a problem for homogeneous cosmology. 
Even though there seems to be a slight preference for CHDM-like models,
the bulk velocity is clearly not the tool for distinguishing between the
variants of our standard picture.
On the other hand, the fact that the model predictions for the bulk
velocity are robust (especially once forced to roughly obey the normalization
constraints from other data) allows
us to use the observed amplitude of $\sim\!300\kms$ bulk velocity on scales
$\lsim 100\hmpc$, in comparison with the observed 
$\delta T/T\sim\!10^{-5}$ in the CMB, as a unique probe of the fluctuation 
growth rate. This provides the most convincing confirmation for our 
basic hypothesis that structure has evolved by gravitational instability 
(see Bertschinger, Gorski \& Dekel 1989; Zaroubi \etal 1997a).

\section{Local Neighborhood}
\label{sec:sbf}

\begin{figure}
\vskip -0.3cm
\centerline{\psfig{file=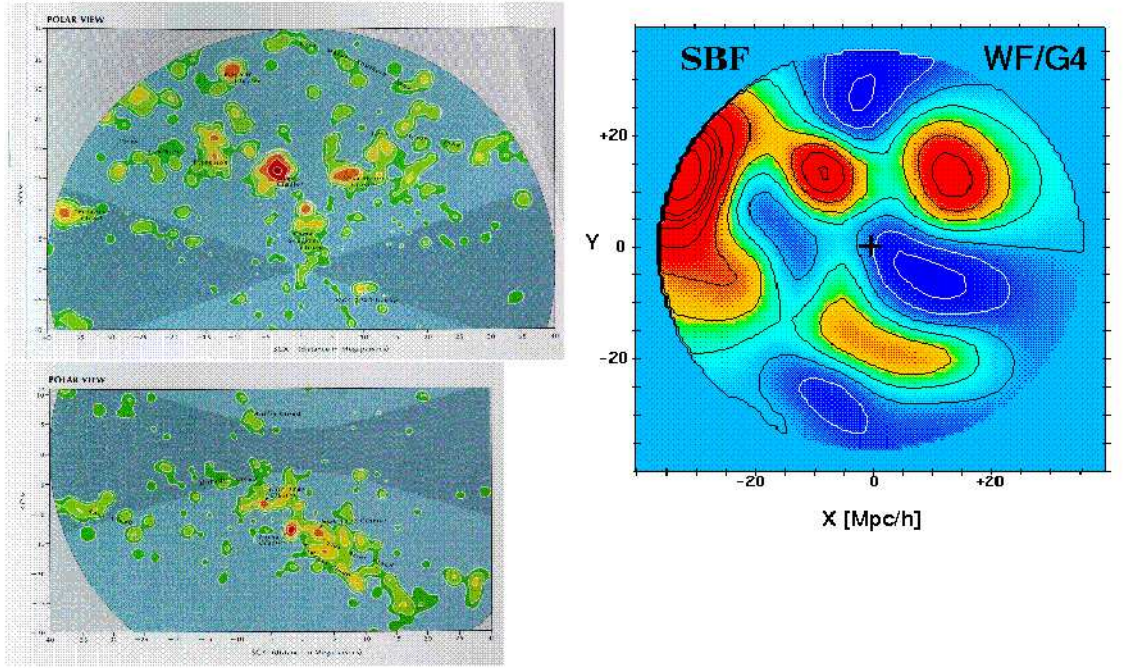,width=15cm}}
\vskip -0.5cm
\caption
{
The local Supergalactic Plane at high resolution.
{\it Right:} Reconstructed mass density field using Wiener Filter
of $4\hmpc$ from SBF peculiar velocities.
{\it Left:} Galaxy distribution from the Nearby Galaxies Atlas
(distances in $h_{75}^{-1} {\rm Mpc}$).
Local clusters such as Virgo (middle), Ursa Major (right) and Fornax (bottom), 
are recovered as attractors.
}
\label{fig:sbf}
\end{figure}

The new SBF peculiar velocities 
(Tonry, Dressler, \tv; Blakeslee \etalns, \tv), in which the distance 
of each galaxy is estimated with unprecedented accuracy 
and Malmquist biases are small, allow a high-resolution study of 
the dynamics in our local cosmological neighborhood, within $30\hmpc$
of the LG. \fig{sbf} demonstrates the potential of this data 
via a high-resolution map of the mass-density field as recovered
by a Wiener-Filter.  This method (Zaroubi, Hoffman \& Dekel 1999;
originally Kaiser \& Stebbins 1991)
provides the most likely mean density field, given the noisy data and an
assumed model for the power spectrum
(in this case a tilted $\Omega=1$ CDM which best fits the M3 data).
The method assumes that both the density fluctuations and the errors
are Gaussian, and it uses linear gravity. Note that the WF induces 
variable smoothing as a function of the local noise;
it allows a high-resolution analysis nearby,
where the data is of high quality,
with an effective smoothing of nearly G4 (a Gaussian window of $4\hmpc$),
compared to $\sim\!$G10 with the M3 and SFI data on larger scales.

While showing (on the left) the near side of the known Great Attractor (GA),
the map reveals for the first time fine dynamical entities nearby. 
The counterparts of these structures in the galaxy distribution are 
clearly seen in the corresponding maps from the Nearby Galaxies Atlas 
(Tully \& Fisher 1987, plates 15 and 19).
For example, the Virgo and Ursa Major clusters,
branching out from the GA along Y$\sim\!15\hmpc$ all the way to X$\sim\!30$,
and the Fornax complex, stretching in the south Galactic hemisphere (Y$<0$)
out to X$\sim 20$. 
The general similarity between the galaxy clusters
and the underlying mass attractors is encouraging.  A quantitative
comparison would allow a study of the non-trivial biasing relation 
between galaxies and dark matter in the local vicinity, on  
scales smaller than addressed so far. 

A sample of $\sim\!500$ TF peculiar velocities within $30\hmpc$
is being completed by Pierce, Tully and coworkers, and the ENEAR 
survey will add
ellipticals in this region.  Together with the accurate SBF data,
they present a new opportunity for high-resolution dynamical analysis of the
local neighborhood.
For example,
these new data call for a revisited VELMOD analysis
of comparison between peculiar velocities and
a whole-sky redshift survey (Willick \etal 1997b; Willick \& Strauss
1999). It should be borne in mind that a proper high-resolution analysis 
must treat nonlinear effects in a reliable way, which must be tested
using proper mock catalogs (\se{mock}). 

\section{Great Attractor and Perseus Pisces}
\label{sec:gapp}

\begin{figure}
\centerline{\psfig{file=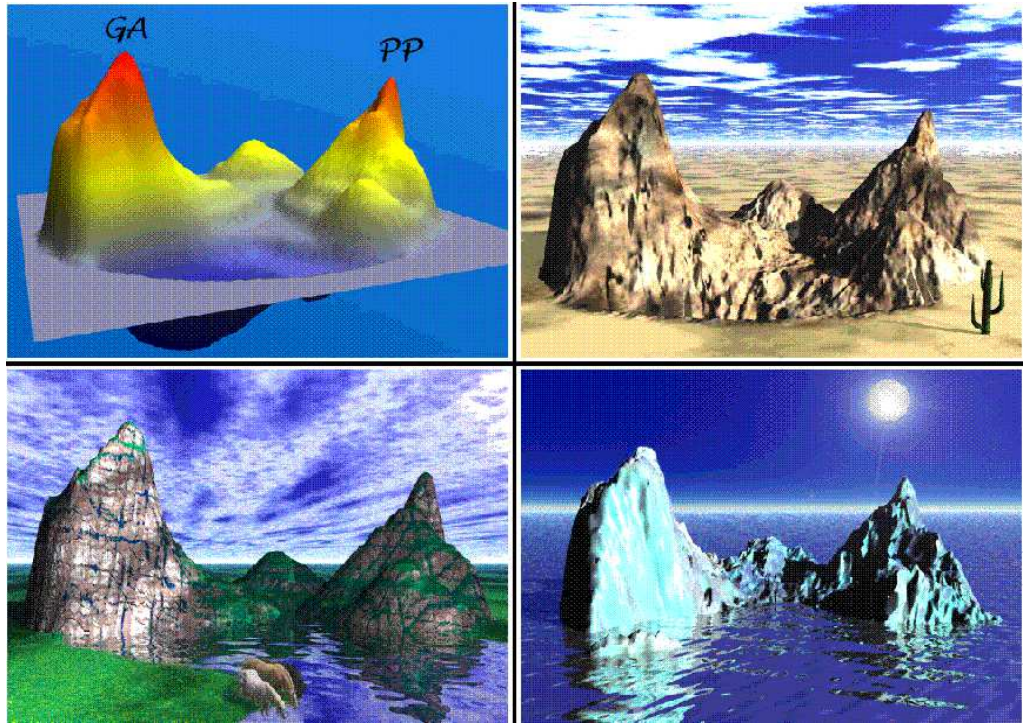,width=13.5cm}}
\caption{Top-left: mass-density distribution in the
Supergalactic plane out to $70\hmpc$, by POTENT G12 from M3
peculiar velocities, featuring the GA, PP, and the big void in between.
The height represents density.
The other panels are artist concepts (courtesy of Ofer Dekel).
The icebergs in the bottom-right reflect the ``coldness" of the flows...
}
\label{fig:four}
\end{figure}

The M3 and SFI datasets, soon to be cross-calibrated with the
whole-sky Shellflow, and then to be complemented by ENEAR, provide 
a rich body of peculiar velocity data for a quantitative analysis of the
dynamical fields on intermediate-large scales.
By applying methods like POTENT to these data we obtain
reliable reconstructions with uniform G12 smoothing out to
$\sim\!60\hmpc$ (at least in several directions). 
\fig{four} shows the G12 density field in the Supergalactic plane 
as extracted by POTENT from the M3 peculiar velocities.
The dominant structures are the Great Attractor (GA, left), Perseus-Pisces
(PP, right), and Coma (back), with the big void stretching in between.

\begin{figure}
\vskip -1cm
\centerline{\psfig{file=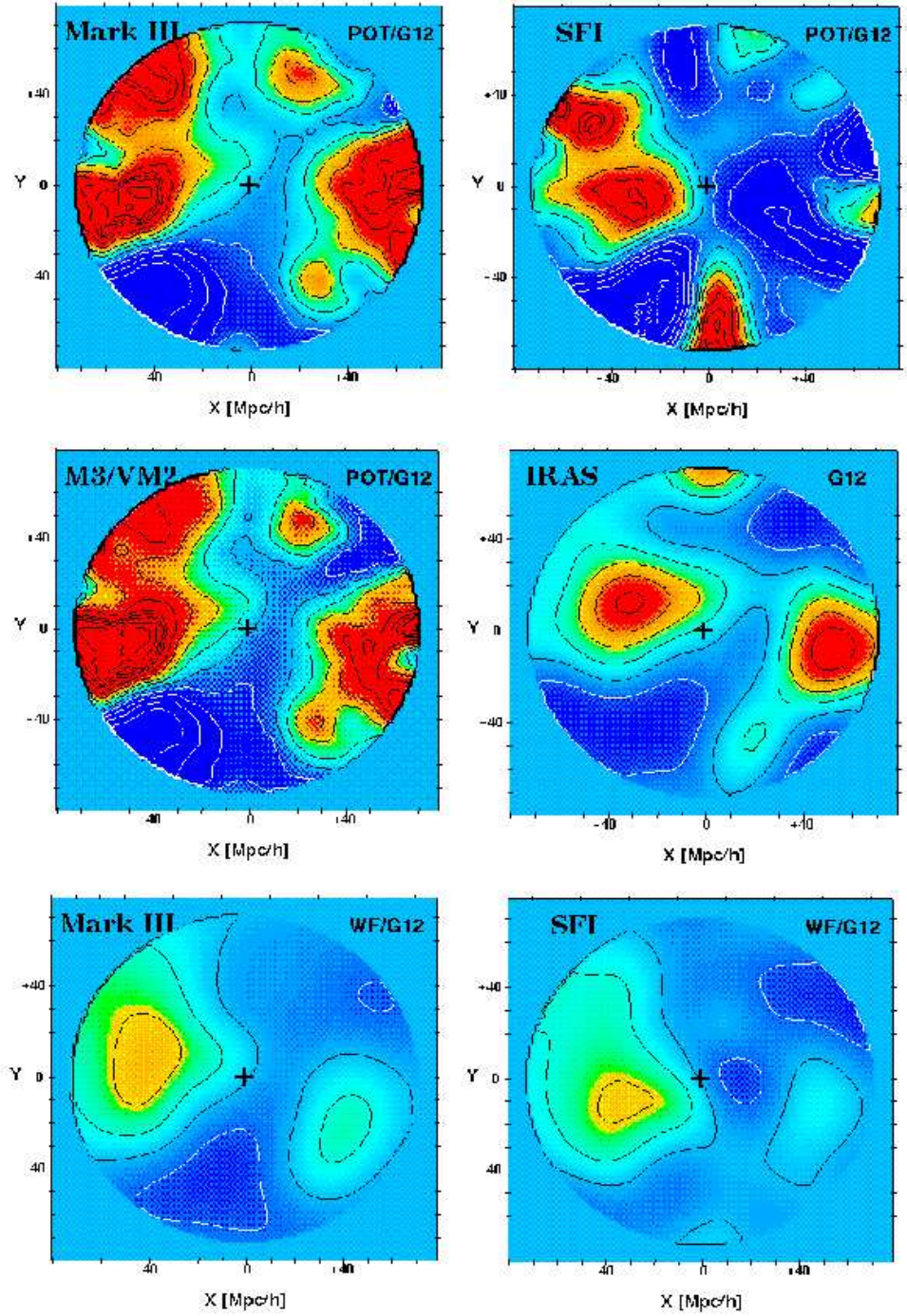,width=13cm}}
\vskip -0.5cm
\caption{
Density maps of the Supergalactic plane out to $70\hmpc$.
The G12 POTENT reconstructions from M3 (original and VM2 calibration,
top- and middle-left)
and SFI peculiar velocities (top-right) can be compared to each other
and to the IRAS galaxy distribution (middle-right).
The WF maps (bottom) highlight the robust features.
}
\label{fig:m6}
\end{figure}

\fig{m6} shows Supergalactic density maps as reconstructed from different
datasets and by different methods. 
The VM2 calibration of M3, which has been tailored to maximize
the agreement of M3 with the IRAS 1.2Jy redshift survey, hardly makes a 
difference to the density map (while it does reduce the bulk flow
somewhat).
The appearance of the GA is quite similar in M3 and SFI, while
PP in SFI is lower and located further away, with the big void  
between the LG and PP deeper and more extended (and thus pointing to
a larger value of $\omm$, \se{cospar}).  
There is a general similarity between the dynamical mass-density maps
(for M3 more than SFI) and the IRAS 1.2Jy galaxy-density map,
allowing a reconstruction of the local biasing field (\se{bias}).
The WF mean field density contrast at a given location is, by construction,
correlated with the quality of the data there. The WF maps thus 
demonstrate that the M3 and SFI densities are similar in the regions 
of high-quality data, such as the GA region, and they highlight the 
robust large-scale dynamical features in our neighborhood.
The M3 and SFI results differ mostly in their bulk velocities in shells
near $50-60\hmpc$ --- a problem that Shellflow may help resolving.

\section{Decomposition: Local and Tidal Components}
\label{sec:deco}

There has been a lot of discussion over the years about
which object is responsible for what velocity.
In general, this discussion is conceptually confused because
the acceleration at a point is the integral of the density fluctuations 
over all of space and it cannot be uniquely assigned to any specific source.  

\begin{figure}[t!]
\vskip -1truecm
\centerline{\psfig{file=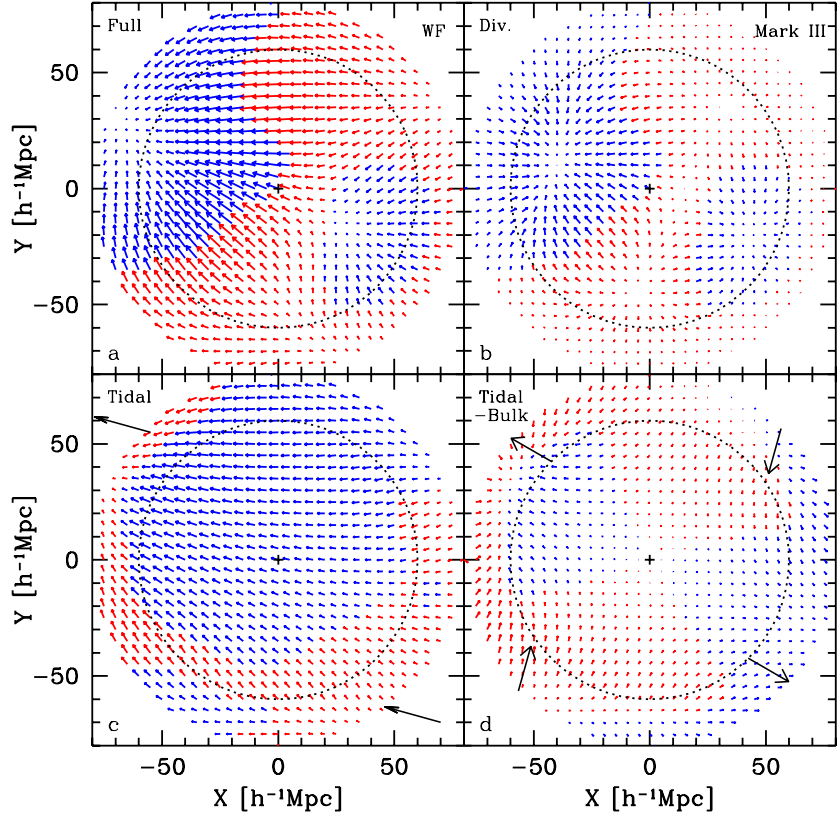,width=12cm}}
\vskip -0.3truecm
\caption{
A decomposition of the M3 WF velocity field in the Supergalactic 
plane (a) into its ``divergent" component, due to the mass fluctuations within
$60\hmpc$ (b), and its ``tidal" component, due to the external mass 
distribution (c).
The residual after subtracting the bulk velocity
from the tidal component is the shear,
including quadrupole and higher moments in the tidal field (d).
Velocities and distances are measured in $100\kms$ or $\hmpc$.
In the upper panels, the color marks the sign of $\divv$, and thus hints
to the local density contrast.
In (c), blue refers to velocities that are aligned ($\pm 30^\circ$) with 
the tidal bulk velocity.
In (d), the color corresponds to the sign of the radial velocity,
highlighting the quadrupole.
}
\label{fig:decomp_wf}
\end{figure}

Nevertheless, given a specific volume, one can uniquely decompose the velocity 
at any point into two well-defined components: a ``divergent" 
and a ``tidal" component, due to the density fluctuations within the volume
and outside it, respectively.
A demonstration of such a decomposition is shown in \fig{decomp_wf},
for the mean velocity field recovered using a WF from
the M3 peculiar velocities, with respect to a sphere
of radius $60\hmpc$ about the LG.
The WF velocity field is first translated into a density field via linear
theory, $\delta \propto\! -\divv$, and then the divergent velocity field
is reconstructed by integrating the inverse of this Poisson equation
inside the sphere of $60\hmpc$. The tidal field is obtained by
subtracting the divergent component from the total velocity.
 
The divergent component shows the main features of convergence and
divergence within the volume, associated with the GA, 
PP, and the voids in between. The CMB velocity of the LG
is about half divergent, namely due to GA, PP and such, and half
tidal, due to mass fluctuations external to the $60\hmpc$ sphere.
There is no significant bulk velocity in the divergent component 
inside that sphere (although there could be one, e.g., if there was
only a single dominant off-center attractor in this sphere);
The bulk velocity inside the sphere of $60\hmpc$ is all tidal, 
due to external fluctuations.
When this bulk velocity is subtracted from the tidal component, one
recovers the shear field, dominated by the quadrupole and
higher moments. The major eigenvector of
the shear tensor lies roughly along the line connecting the LG with
the Shapley supercluster. A fit of a simplified toy model made of a single 
point-mass attractor to the tidal component yields a mass excess of
$\approx 4 \times 10^{17} h^{-1}\Omega ^{0.4} M_\odot$
at a distance of $\approx 175\hmpc$ in the direction of Shapley.

This analysis thus allows us to extract information from the velocities in
a given volume about the mass distribution outside this volume, and it
can be applied to different datasets and different volumes. 
For example,
when applied to the WF (or POTENT) velocities from the SFI data inside
$60\hmpc$, the tidal bulk velocity turns out smaller than in the M3 case, 
but the residual shear field is very similar, indicating a similar 
quadrupole and external sources.
When applied to the SBF data within $30\hmpc$,
the decomposition yields similarly that the bulk velocity is dominated 
by the tidal field, and the major axis of the shear tensor lies roughly
along the line connecting PP, LG and GA.

\section{Very Large Scales}
\label{sec:vls}

\begin{figure}
\vskip -0.5cm
\centerline{\psfig{file=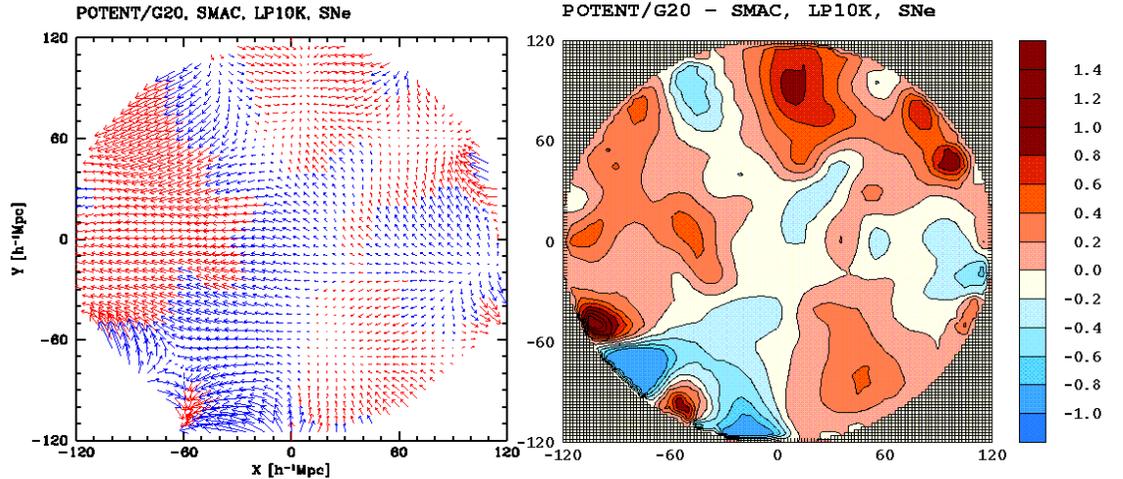,width=15cm}}
\vskip -0.3cm
\caption{
G20 POTENT reconstruction on very large scales from peculiar
velocities of clusters and SNe.
The near sides of the big mass enhancements associated
with the Shapley and Horologium superclusters show up on the left, 
and with Coma at the top.
}
\label{fig:smac}
\end{figure}

The new data of peculiar velocities for clusters of galaxies on large
scales allow dynamical reconstruction beyond just the 
bulk velocity. As a demonstration, \fig{smac} shows G20 POTENT maps
extracted from a combination of the 
SMAC, LP10K, and SN data out to $120\hmpc$.
Beyond the familiar structures of GA and PP that dominate the
inner $\sim\!60\hmpc$, one can see the Coma structure at Y$\sim\!50-100\hmpc$,
and the near sides of the Shapely (Y$>0$) and Horologium (Y$<0$)
overdensities behind the GA, at X$\sim\!-100\hmpc$ and beyond.
The earlier hints from the tidal component of the velocities at smaller
distances (\fig{decomp_wf}) are now beginning to be confirmed by the 
local derivatives of the peculiar velocities directly measured at 
large distances.

Another example for a large-scale study is the monopole analysis, which could 
constrain a local Hubble Bubble (Zehavi \etal 1997; Dale \& Giovanelli,
\tv; Fruchter, \tv) and thus modify the local estimates of $h$ and $\omm$ 
(\se{cospar}). 
Peculiar velocities of many objects both inside the void and far
outside it are necessary for a reliable result.
With more and more data at large distances, the monopole deviations
from the universal Hubble flow could be determined with increasing
accuracy, because the error $\delta H=\delta v/r$ is 
independent of distance (as $\delta v \propto r$).

I think that the greatest potential for future studies of
local cosmic flows lies in big surveys of SN Ia (Riess, \tv). They provide 
a distance indicator of only 5-10\% error which can be
observed out to hundreds of megaparsecs and is, in principle,
of unlimited sampling density, limited in practice only by the patience
and dedication of the observers.

\section{Cosmological Parameters}
\label{sec:cospar}

I share the discontent expressed by Strauss (\tv) and others 
from the fact that the results from cosmic flows have been 
unjustifiably underrated by many in the community.
This has a lot to do with bad PR on our side, where in many cases
we tend to stress marginal apparent discrepancies between different results
and take for granted the robust valuable findings.

An important feature of peculiar velocity data is that it allow us to
addresses directly the dynamics of the total (cold) mass distribution, 
and thus bypass the difficulties introduced by density biasing of the luminous
galaxies, which are unavoidable in the analysis of redshift surveys. 
For example, the spatial velocity variations provide
direct constraints on the value of the cosmological density parameter
$\omm$.
This makes them valuable even when the errors are not yet as small and
under control as they could be, given that the available complementary 
data all involve additional parameters such as $\sigma_8$, $\oml$, or 
biasing parameters, and they all have their own appreciable errors.
The results obtained on intermediate scales directly from M3 and SFI 
constrain $\omm$ at the $\pm 2\sigma$ level to the range 0.3-1.0
(Primack, \tv; based on
Nusser \& Dekel 1993; Dekel \& Rees 1993; Bernardeau \etal 1995;
and yet unpublished results from the newer data).

Allowing the power spectrum to be of the CDM type, 
the maximum-likelihood estimate of $\omm$ is $0.5 \pm 0.1$ 
(Zaroubi \etal 1997b; Freudling \etal 1999).
A similar analysis with more free parameters can constrain 
additional parameters that affect the power spectrum, such as
the large-scale power index $n$, or the normalization parameter
$\sigma_8$.
 
The results from cosmic flows provide valuable orthogonal constraints 
to complementary data.  
For example, combined with the constraints from the global
geometry of space-time based on high-redshift supernovae type Ia,
which is roughly $0.8\omm-0.6\oml=-0.2\pm0.1$
(Riess \etal 1998; Perlmutter \etal 1999),
the velocity constraints on $\omm$ confine the value of $\oml$
to $0.8\pm0.3$ (Zehavi \& Dekel 1999).
Jointly with available CMB constraints as well, one can obtain simultaneous
constraints on three parameters, such as $\omm$, $\sigma_8$ and $h$ 
(Zehavi \& Dekel, \tv), still without appealing to biasing parameters.
The addition of constraints from the abundance of clusters
(Eke, Cole \& Frenk 1996), or from gravitational lensing,
should allow us to confine these dynamical parameters with
even higher accuracy.

We heard evidence for the ``coldness" of the local flow (Lake, \tv;
Van de Weygaert \& Hoffman, \tv; Klypin, \tv),
which left us still wondering whether it is really in conflict with
standard models (Strauss, \tv).
I dare to report on a very preliminary result of a likelihood analysis
based on M3 and SFI (extending Zehavi \& Dekel, \tv), 
which seems to favor a power spectrum that drops 
sharply at $k\gsim k_{\rm peak}$.  Such a power spectrum could be obtained, 
for example, with a high faction of baryonic or hot dark matter. 
A similar, independent hint comes from the SMAC data (Hudson \etalns, \tv).
This would add to the uncertainty of the results obtained under the
assumption of $\Lambda$CDM.

When a redshift survey is involved, the unknown biasing relation
between galaxies and mass introduces another source of uncertainty,
which should not be ignored. One should treat biasing properly
before $\omm$ can be extracted from the range of estimates
of parameters like $\beta$ (\se{bias}).

Many different clever ideas of how to estimate cosmological parameters from 
peculiar velocities can be thought of. Some turn out to be more 
discriminatory and less biased than others. A given idea can turn into
a viable method, whose results should be considered seriously,
only after the method has
been tested and calibrated using proper mock catalogs,
and a detailed error analysis is provided (\se{mock}).
If this attitude is adopted by all practitioners, the field will 
regain the respectability it deserves in evaluating the cosmological
parameters.

\section{Biasing}
\label{sec:bias}

\begin{figure}
\vskip -0.3cm
\hskip -0.3cm
\centerline{\psfig{file=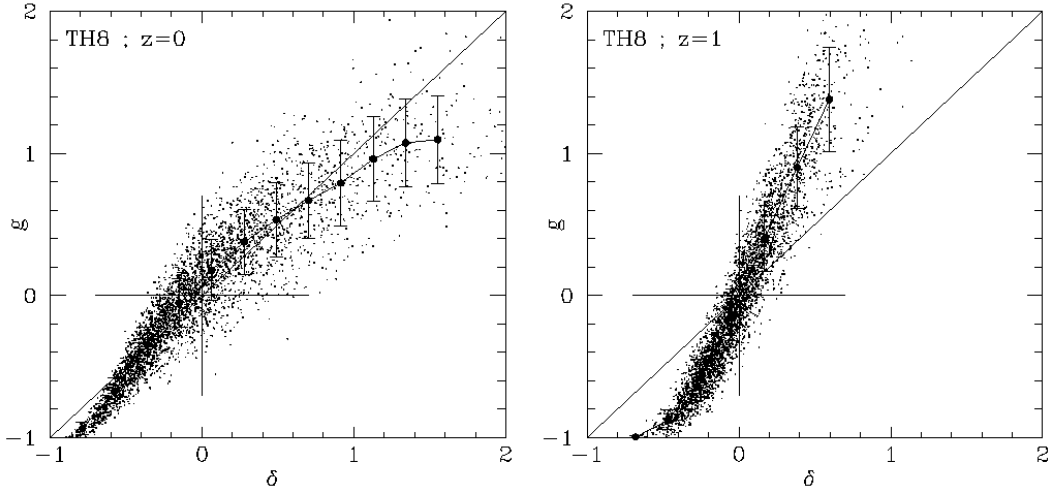,width=14cm}}
\vskip -0.3cm
\caption{Biasing of galactic halos ($>\! 2 \times 10^{12} \msun\!$)
versus mass in an $N$-body simulation,
demonstrating nonlinearity and stochasticity.
The smoothed (top-hat $8\hmpc$) density fluctuation fields are plotted 
at grid points.
The conditional mean and scatter are marked (bars).
{\it Left:} at $z=0$, when $\sigma_8=0.6$, showing characteristic
nonlinear behavior in underdense regions.
{\it Right:} at $z=1$, when $\sigma_8=0.3$, demonstrating strong time
evolution.
}
\label{fig:b01}
\end{figure}

Understanding the biasing relation between galaxies and mass is crucial
for the purpose of translating measurements of bias-contaminated 
quantities such as
$\beta$ (vaguely defined as $\Omega^{0.6}/b$) into accurate estimates 
of $\Omega$.  On the other hand, the biasing can provide hints for the 
complicated physical processes involved in galaxy formation. 

The linear deterministic relation between the density fluctuations
of galaxies and mass, $\delg(\vx) = b\,\del(\vx)$, has no theoretical
basis and is not self-consistent. Indeed, the analytic analysis of halo
biasing (Mo \& White 1996) predicts that the biasing is non-linear, 
$b=b(\del)$.  Then, the biasing at any other smoothing 
scale must obey a different 
$b(\del)$ and be non-deterministic, i.e., it involves
scale-dependence and scatter.  In addition to shot noise, an inevitable 
source of scatter are hidden variables affecting
the efficiency of galaxy formation beyond its dependence on $\del$,
which are yet to be studied in detail (e.g., Blanton \etal 1998).
\fig{b01} demonstrates some of the nontrivial qualitative features of
halo biasing in $N$-body simulations.  
The nonlinear shape of $b(\del)$ at $\del<0$ is robust,
while at $\del >0$ it varies with mass, time and scale
(see also Frenk, \tv; Sheth, \tv).  

In order to properly incorporate the biasing in the
analysis of cosmic flows, one needs an appropriate formalism that
quantifies non-trivial biasing. For example, in the formalism of 
Dekel \& Lahav (1999), the linear and deterministic
relation at a given scale and time is replaced by the conditional
distribution $P(\delg\vert\delta)$.  The mean nonlinear biasing is 
characterized by the conditional mean
$\av{\delg\vert\delta}\!\equiv\!b(\delta)\,\delta$, and the scatter
by the conditional variance $\sb^2(\delta)$.  To second order, the biasing
is then defined by 3 parameters: the slope $\bh$ of the
regression of $\delg$ on $\delta$ (replacing $b$), a non-linearity
parameter $\bt/\bh$, and a scatter parameter $\sb/\bh$. 
The ratio of variances $\bv^2$ and the
correlation coefficient $r$ mix these fundamental parameters.
In the case shown in \fig{b01} at $z=0$, the overall non-linearity 
is $\bt^2/\bh^2=1.08$, and the scatter is $\sb^2/\bh^2=0.15$.
These effects lead to differences of order 20-30\%
among the various measures of $\beta$. 
An additional contribution to the span of $\beta$ estimates
may have to do with the biasing dependences on scale and on galaxy properties.
These are deduced from simulations such as the one shown in \fig{mock} 
(e.g., Somerville \etal 1999), and can be measured from
large redshift surveys with type identification, such as SDSS.
Together, these nontrivial biasing features could 
explain much of the observational range, $\betai=0.4-1.0$ (Strauss, \tv,
Table 2).  Any outliers are suspicious of underestimated errors,
and should be re-examined using proper mock catalogs (\se{mock}).

\section{Error Analysis}
\label{sec:mock}

The research field of cosmic flows, which started in the eighties with
semi-qualitative analyses, has developed into a mature, quantitative phase 
in which the errors ought to be evaluated in detail.  
This will allow us to understand the range of
estimates for parameters like $\beta$ and $\omm$.
Measurements based on new methods or data which are not accompanied by 
a detailed error analysis are not very useful at this point
(though such results are still being presented at times). 

An appropriate tool for error analysis is an ensemble of Monte
Carlo mock catalogs, in which both the nonlinear gravitational dynamics and 
the galaxy formation process are simulated properly, and then galaxies 
are sampled and measured in a way that mimics the observational
procedure.  Such mock catalogs allow an evaluation of both random
and systematic errors.

The development of the POTENT method (Dekel \etal 1999; Kolatt, \tv) 
is an example. The recovery algorithm and the associated
methods for measuring cosmological parameters
have been calibrated based on mock catalogs by Kolatt \etal (1996).
A key feature of these simulations was the effort to mimic the actual 
structure in our local neighborhood, generating the initial conditions
using constrained realizations 
based on the density of galaxies in the IRAS 1.2Jy redshift survey.
Such simulations allow for correlations between the errors and the
underlying density field.
The main limitations of these mock catalogs were the 
unsatisfactory treatment of nonlinear effects
due to limited resolution in the simulations, the simplified way of
identifying galaxies, and the fact that the
simulations were initially restricted to an $\omm=1$ standard CDM cosmology.
It is now time for a new generation of mock catalogs that will overcome
these limitations.

\begin{figure}
\vskip -0.3cm
\hskip -0.3cm
\centerline{\psfig{file=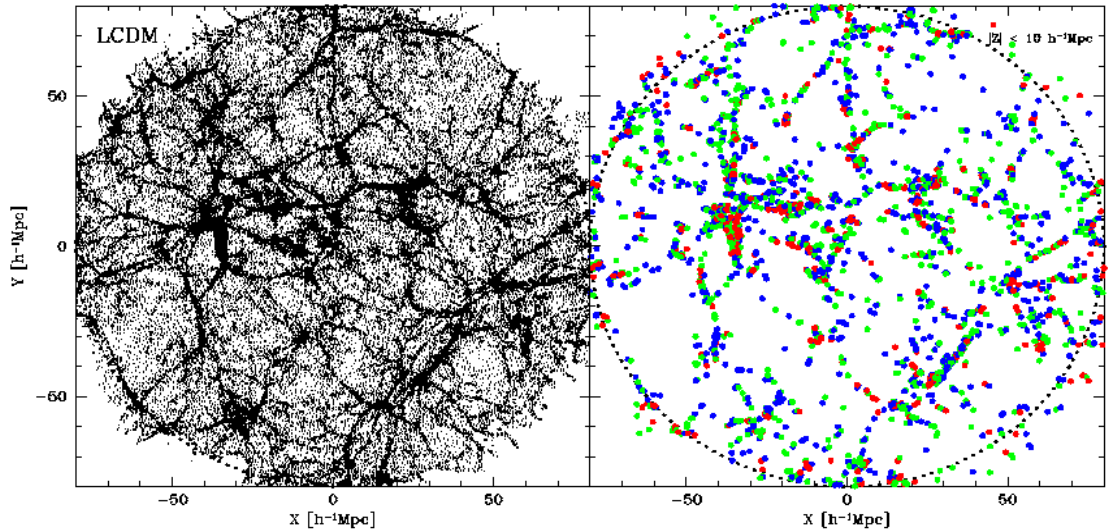,width=15cm}}
\vskip -0.3cm
\caption{
Constrained simulations in the Supergalactic Plane.
The large-scale structure is constrained to fit the density in the
IRAS 1.2Jy redshift survey, while the small-scale fluctuations
are drawn from a $\Lambda$CDM power spectrum.
{\it Left:} Dark Matter, showing the GA (top-left),
PP (bottom-right), etc.
{\it Right:} Modeled galaxies in the same slice, color coded by B-V,
demonstrating how red, early-type galaxies reside in clusters.
This is the basis for mock catalogs that serve as benchmarks for
methods.
}
\label{fig:mock}
\end{figure}

New mock catalogs of this sort are becoming available, based on the
GIF simulations (Eldar \etal 1999).
Constrained realizations (based on IRAS 1.2Jy) serve as initial conditions
that were evolved forward in time using a high-resolution parallel tree code, 
assuming either $\tau$CDM ($\omm=1$) or $\Lambda$CDM ($\omm=0.3$),
both with power spectra that allow
a simultaneous fit to COBE normalization and the observed cluster abundance.
Particle positions and velocities were stored at 50 logarithmically
spaced time-steps in order for different recipes of galaxy formation
to be implemented {\it post hoc} in considerable detail.
The physical processes include shock heating (and possibly radiative
heating) of the pre-galactic gas, radiative cooling, star-formation, 
hydrodynamic (and possibly radiative) feedback from supernovae, and 
enrichment with heavy elements.
\fig{mock} shows a slice from the $\Lambda$CDM constrained simulation,
comparing the dark-matter distribution with the galaxy distribution.

Given the relevant properties for each of the galaxies, such as
magnitude and internal velocity, one can ``observe" a set
of Monte-Carlo mock catalogs following the selection criteria 
and specifications of each of the observed catalogs (e.g., Diaferio
\etal 1999).
By applying one's algorithm to these mock catalogs, for which the 
``true" underlying dynamics is known,  one can quantify
the random and systematic errors in detail.
These simulations and mock catalogs will soon be made available 
as standard benchmarks for reconstruction methods.
Designer mock catalogs for specific new datasets can be made to order.

\section{Conclusion}
\label{sec:conc}

\def\nb{\noindent$\bullet$\ }

My main points are as follows:

\nb
The observed amplitudes of bulk velocity out to $\sim\!100\hmpc$
(with the marginal exception of the current BCG result)  
are consistent with our standard family of cosmological models.
Full ``convergence" is not really required on any scale.
The main lesson from the bulk velocity is its general consistency with the
gravitational growth rate of perturbations starting from the 
fluctuations at recombination as measured in the CMB.

\nb
The SBF and other data provide an opportunity for high-resolution 
dynamical analysis of the local neighborhood out to $30\hmpc$. 
Virgo, Ursa Major and Fornax show up as local attractors,
and can help us model the biasing relation on small scales,
provided that nonlinear effects are treated properly.

\nb
The dynamical structure of the GA is robust in the M3 and SFI datasets.
The Shellflow data should improve the cross-calibration of North
and South in the M3 and SFI data, which, together with other data,
ought to allow an accurate evaluation of the bulk velocity out 
to $\sim\!70\hmpc$.

\nb
A decomposition of the velocity field into divergent and tidal
components allow us to tell that a significant part of the bulk
velocity inside $60\hmpc$ is due to external density fluctuations,
and that the shear field points at the Shapely concentration as 
a massive attractor.
 
\nb
The extended cluster and SN velocities enable dynamical analysis
beyond just bulk flow out to $120\hmpc$, confirming the mass
enhancements associated with Coma, Shapely and Horologium.
The available data provide marginal evidence for a local Hubble Bubble,
that should become less ambiguous with more SN data.

\nb
Supernovae type Ia seem to be the most promising tool for cosmic flow analysis.
The SN hunters are thus encouraged to pursue large surveys at low redshifts.

\nb
Peculiar velocities do provide interesting constraints on cosmological
parameters.  For example, they confine $\omm$ to the range 0.3-1.0
at 95\% confidence, independent of biasing or other data, solely
based on the assumption of Gaussian initial fluctuations.
Combined with other data, this constraint is translated to 
constraints on other parameters, such at $\oml$, $\sigma_8$, $h$, etc.

\nb
Galaxy biasing is an obstacle for translating a measured value of $\beta$ into
an estimate of $\omm$. Nontrivial features of the biasing scheme, 
including nonlinearity, stochasticity, scale dependence and type
dependence, as predicted by models and simulations, can explain
much of the span of estimates for $\beta$.

\nb
Quantitative error analysis is essential in order to complete the
transition of large-scale dynamics into a mature field.
Every method has to be calibrated with appropriate mock catalogs,
in which nonlinear dynamics and galaxy formation are simulated properly.
Such mock catalogs are being produced and offered as benchmarks.

\smallskip
Where next?
The field of cosmic flows enjoyed several influential conferences,
starting in Hawaii, Rio and the Vatican in 1985-1987, then Paris in
1993, and now Victoria in 1999.  Projecting ahead,
we should look forward to meeting again with exciting new results
around 2005. This is provided that somebody energetic like 
Stephane Courteau takes charge in organizing such a conference.

\acknowledgments

I am especially indebted to Ami Eldar, the current guardian of POTENT,
for the computations and maps.
I am grateful to all my collaborators, many of which have participated in
this conference. 
Our work has been partially supported by the  
US-Israel Binational Science Foundation (95-00330, 98-00217),
by the Israel Science Foundation (950/95, 546/98), and
by NASA (ATP NAG 5-301).
I believe I represent all the conference participants in thanking 
Stephane Courteau for organizing this very successful conference.

\def\re{\reference}

\vfill\noindent
{To appear in ``Cosmic Flows:
Towards an Understanding of Large-Scale Structure",
eds S. Courteau, M.A. Strauss, \& J.A. Willick,
ASP Conf. Series}


\begin{references}

\re Bernardeau, F., Juszkiewicz, R., Dekel, A., \& Bouchet, F.R.
    1995, MNRAS, 274, 20
\re Bertschinger, E., Gorski, K., \& Dekel, A. 1990, Nature, 345, 507
\re Dale, D.A., Giovanelli, R. , Haynes, M.P., Campusano, L.E.,
     Hardy, E., \& Borgani, S. 1999, ApJL, 510, L11 (SCI+II)
\re Blanton, M., Cen, R., Ostriker, J.P., \& Strauss, M.A.
    1999, ApJ, 522, 590
\re Dekel, A., Burstein, D., \& White, S.D.M. 1997, in Critical 
    Dialogues in Cosmology, ed. N. Turok (World Scientific), p. 175
\re Dekel, A., Eldar, A., Kolatt, T., Yahil, A., Willick, J.A., 
    Faber, S.M., Corteau, S., \& Burstein, D. 1999, ApJ, 522, 1 (POTENT)
\re Dekel, A., \& Lahav, O. 1999, ApJ, 520, 24
\re Dekel, A., \& Rees, M.J. 1993, ApJL, 422, L1
\re Diaferio, A., Kauffmann, G., Colberg, J.M., \& White, S.D.M. 1999,
    MNRAS, 307, 537
\re Eldar, A., Dekel, A., Lemson, G., Mathis, H., Kauffmann, G., 
    \& White, S.D.M. 1999, in preparation
\re Eke, V.R., Cole, S., \& Frenk, C.S. 1996, \mnras, 282, 263
\re Freudling, W., Zehavi, I., daCosta, L.N., Dekel, A., Eldar, A.,
    Giovanelli, R., Haynes, M.P., Salzer, J.J., Wagner, G., \& Zaroubi, S.
    1999, \apj, 523, 1
\re Haynes, M.P., Giovanelli, R., Salzer, J.J., Wegner, G., Freudling,
    W., da Costa, L.N., Herter, T., \& Vogt, N.P. 1998a, AJ, 117, 1668 (SFI)
\re Haynes, M.P., Giovanelli, R., Chamaraux, P., da Costa, L.N.,
    Freudling, W., Salzer, J.J., \& Wegner, G. 1998b, AJ, 117, 2039 (SFI)
\re Hudson, M.J., Smith, R.J., Lucey, J.R., Schlegel, D.J. \&
    Davies, R.L. 1999, ApJL, 512, L79  (SMAC)
\re Kaiser, N. 1988, \mnras, 231, 149
\re Kaiser, N. \& Stebbins, A. 1991, in Large Scale Structure
    and Peculiar Motions in the Universe, eds. D.W. Latham \& L.N. da
    Costa (ASP Conference Series), p. 111
\re Kolatt, T., Dekel, A., Ganon, G., \& Willick, J. 1996, \apj, 458, 419
\re Lauer, T.R. \& Postman, M. 1994, \apj, 425, 418 (BCG)
\re Mo, H.J. \& White, S.D.M. 1996, MNRAS, 283, 154
\re Nusser, A., \& Dekel, A. 1993, \apj, 405, 437
\re Perlmutter, S. \etal 1999, \apj{517}, 565
\re Riess, A.G., Davis, M., Baker, J., \& Kirshner, R.P., 1997, 
    ApJL, 488, L1 (SNIa)
\re Riess, A.G. \etal 1998, AJ, 116, 1009
\re Somerville, R.S., Lemson, G., Sigad, Y., Dekel, A., Kauffmann, G.,
    \& White, S.D.M. 1999, MNRAS, submitted
\re Tully, R.B., \& Fisher, J.R. 1987, Nearby Galaxies Atlas (Cambridge
    University Press)
\re Watkins, R. \& Feldman, H.A. 1995, ApJL, 453, L73
\re Willick, J.A. 1999, \apj, 516, 47 (LP10K)
\re Willick, J.A., Courteau, S., Faber, S.M., Burstein, D., Dekel, A., \&
    Strauss, M.A. 1997a, ApJS, 109, 333 (M3)
\re Willick, J.A., Strauss, M.A., Dekel, A., \& Kolatt, T. 
    1997b, \apj, 486, 629 (VELMOD)
\re Zehavi, I., \& Dekel, A. 1999, Nature, 401, 252
\re Zehavi, I., Riess, A.G., Kirshner, R.P., \& Dekel, A. 1998, \apj, 503, 483
\re Zaroubi, S., Sugiyama, N., Silk, J., Hoffman, Y., \& Dekel, A. 
    1997a, \apj, 490, 473
\re Zaroubi, S., Zehavi, I., Dekel, A., Hoffman, Y. \& Kolatt, T. 1997b, 
    \apj, 486, 21
\re Zaroubi, S., Hoffman, Y., \& Dekel, A.  1999, \apj, 520, 413
\re Zehavi, I., Riess, A.G., Kirshner, R.P., \& Dekel, A. 1998, \apj, 503, 483.

\end{references}
\end{document}